\documentclass[final,3p,times,twocolumn]{elsarticle}

\usepackage{soul}
\setstcolor{red}
\usepackage{graphicx}
\usepackage{amssymb}

\usepackage{bm}

\usepackage{siunitx}

\usepackage{xcolor,cancel}

\usepackage{lineno}

\usepackage{color}

\usepackage{verbatim}

\usepackage[font=small,labelfont=bf,margin=\parindent,tableposition=top]{caption}

\begin{document}

\begin{frontmatter}


\title{A polymer based phononic crystal}


\author[ceb]{Nan~Li\corref{cor}}
\ead{nl276@cam.ac.uk}
\cortext[cor]{Corresponding author}
\author[cat]{Christopher R.~Lowe}
\author[ceb]{Adrian C.~Stevenson}
\address[ceb]{Department of Chemical Engineering and Biotechnology, University of Cambridge, Philippa Fawcett Drive, Cambridge, CB3 0AS, UK}
\address[cat]{Cambridge Academy of Therapeutic Sciences, University of Cambridge, 17 Mill Lane, Cambridge, CB2 1RX, UK}

\begin{abstract}
A versatile system to construct polymeric phononic crystals by using ultrasound is described. In order to fabricate this material a customised cavity device fitted with a $\sim$2 MHz acoustic transducer and an acoustic reflector is employed for standing wave creation in the device chamber. The polymer crystal is formed when the standing waves are created during the polymerisation process.  The resulting crystals are reproduced in the shape of the tunable cavity device, and add unique periodic features. Their separation is related to the applied acoustic wave frequency during the fabrication process and their composition was found to be made up to two material phases. To assess the acoustic properties of the polymer crystals their average acoustic velocity is measured relative to monomer solutions of different concentrations.  It is demonstrated that one of the signature characteristics of phononic crystal, the slow wave effect, was expressed by the polymer. Furthermore the thickness of a unit cell is analysed from images obtained from an optical microscope. By knowing the thickness the average acoustic velocity is calculated to be 1538 m/s when the monomer/cross-linker concentration is 1.5 M. This numerical calculation closely agrees with the predicted value for this monomer/cross-linker concentration of 1536 m/s. This work provides a methodology for accessing a new type of adaptable phononic crystal based on flexible polymers.
\end{abstract}


\end{frontmatter}


\section{Introduction}
\label{S:intro}
\par Phononic crystals occur naturally as the result of the periodic nature of atomic crystals. The concept of artificial phononic crystals was proposed decades ago \cite{brillouin1953propagation, kushwaha1993acoustic} and significant interest has followed since but not limited \cite{yang2004focusing, zhang2004negative, qiu2005directional, qiu2006acoustic, fang2006ultrasonic, feng2006acoustic, cheng2008broad}. The attraction is that these materials can potentially address enduring engineering challenges in acoustics as they introduce physical effects that change the very nature of acoustic wave excitation and propagation. They can be commonly achieved by periodically altering the density or bulk modulus, so that the acoustic waves propagation depends on wavelength, which in turn alters the group velocity, phase velocity and the non-linear properties of the material. Alternatively for subwavelength periodicity evanescent acoustic surface waves have also been employed \cite{wu2004surface}. Together the breadth of phononic crystal applications is considerable with band gap and band edge states able to disperse the group velocity and bend acoustic waves \cite{schriemer1997energy, page1996group}, leading to applications including signal processing \cite{benchabane2006evidence, mohammadi2011chip} and opportunities to reduce thermal conductivity \cite{gillet2009atomic, hopkins2010reduction}.
\par This rapidly growing  field of study originated in bulk materials, however it was difficult to produce the periodic properties. Initially this was achieved by combining bulk acoustic wave devices to make filters. This activity became easier with surface acoustic wave devices and their metallic grating \cite{wu2004surface, laude2005full}. These have provided excellent acoustic propagation control, and so these devices control acoustic waves to provide signal processing for mobile phones. A key development is the interdigitated electrode structure. A periodic metal pattern is formed on a substrate, and interacts with a surface wave, typically a Rayleigh wave. Unfortunately this method cannot be used for bulk materials, so it cannot be used to modulate sound transmission in the larger environment. Nevertheless some attempts have been made to create bulk phononic crystals as a first step to creating acoustic metamaterials. As a newly emerging field metamaterials display counter-intuitive physical effects. Due to the Bragg scattering caused by impedance contrast of the mass density or the elastic moduli, useful acoustic dispersion, band gap and slow wave effect emerge. Although the theoretical research of phononic crystals is rich in numerical models, fabrications methods are rare. These broadly consist of solid sphere arrays embedded in soft matrix \cite{liu2000locally, sainidou2006locally, cheng2006observation, leduc2016magnetic}. Or more recently, 3D printing technology is used for phononic crystal fabrication \cite{lucklum2015rapid}. Because of these limited options, most of these crystals are built slowly, layer by layer.
\par We report a versatile method to add multiple periodic features using a monomer/cross-linker solution as a starting point. This method applies acoustic standing waves to an acrylamide system undergoing a polymerisation process. In the following sections we consider the theory of phononic crystals alongside bulk grating approaches. We describe the standing wave imprinting device, formulation of monomer/cross-linker mix and go on to analyse polymer images, imprinting mechanism and acoustic transmission properties associated with these polymeric phononic crystals.

\section{Theory}
\label{S:theory}
\par The applications of acoustics often involve the interaction of waves at the acoustic boundaries. Acoustic reflection occurs along a path associated with the incident wave. Only part of the incident wave energy transmits from the first medium, referred as medium \(m\) here, into the second medium, referred as reflector \(r\) here. The most basic acoustic properties that determine this include the medium elasticity and density. The elasticity and density of a medium determine the acoustic impedance \(Z\), which in turn governs the transmission coefficient \(T\) and the reflection coefficient \(R\). Both  \(T\) and \(R\) are independent from the energy flow direction.  Under the condition of normal incidence, 
\begin{equation}
\label{eq:acoustic_R}
R=1-T=\bigg| \frac{Z_r-Z_m}{Z_r+Z_m}\bigg|
\end{equation}
\par A typical acoustic phononic crystal is a collective cluster of a number of unit cells made up of two layers with an acoustic boundary in between. To comprehend the waves propagation in a periodic structure, it is essential to understand the nature of the wave eigenmode in it. Bloch indicated that a wave propagating in a periodic structure is a superposition of a series of plane wave \cite{bloch1929quantenmechanik}. The Bloch theorem is expressed as:
\begin{equation}
\bm{\psi}_k(\bm{r})=\bm{u}_k(\bm{r})e^{i\bm{kr} }
\end{equation}
\par Where $\bm{k}$ is the wave vector, $\bm{r}$ is the position, \(e\) is Euler's number, \(i\) is the imaginary unit. $\bm{u}_k(\bm{r})$ is the periodic function of the crystal lattice with $\bm{u}_k(\bm{r})=\bm{u}_k(\bm{r}+\bm{R})$ in which $\bm{R}$ is the periodicity of the crystal lattice. Being different from the velocity definition of a particle, waves have three types of velocities, including phase velocity \(\bm{v}_{ph}\) , group velocity \(\bm{v}_g\) and energy velocity \(\bm{v}_e\), amongst which the phase velocity refers to the propagation of an equiphase surface; the group velocity to the propagation of a wave pocket; and the energy  velocity to the propagation of energy. The group velocity is proved to be equivalent to the energy velocity of a Bloch wave \cite{sakoda2004optical}, \textit{i.e.}:
\begin{equation}
\bm{v}_g=\frac{\partial{\omega(\bm{k}_r)}}{\partial{\bm{k}_r}}=\frac{\bm{s}_k}{w_k}=\bm{v}_e
\end{equation}
\par Where $\omega_r(\bm{k})$ holds dispersion relation with $\bm{k}_r$, $\bm{s}_k=\langle \bm{s} \rangle|_{\omega(\bm{k}_r)}$ and $w_k=\langle w\rangle|_{\omega(\bm{k}_r)}$ are the average wave energy flow density and wave energy density respectively. While the phase velocity applies to a wave at single frequency, \textit{i.e.}$v_{ph}={\omega}/{k}$.
\par A key feature is when the length scale of the periodic structure is comparable to the wavelength of the wave, the excitation of the unit cells leads to a strong resonant scattering. Thus the action of the phononic crystal can be evidenced by this strong resonant scattering and a "slow wave effect" \cite{schriemer1997energy, page1996group}. 

\section{Experimental setup}
\label{S:setup}
\par In order to fabricate the polymer crystal, a bespoke ultrasonic cavity chamber is employed for holding the monomer/cross-linker solution within the field of defined acoustic standing waves. To assess the subsequent interaction of developed polymer crystals with acoustic waves, a series of average acoustic velocity measurements have been obtained at room temperature. The following subsections present the design concept of this adjustable cavity device, the procedure for fabricating the polymeric phononic crystal and the acoustic velocity measurements needed to characterise the material.

\subsection{Adjustable polymeric phononic crystal fabrication tube}
\label{sS:PC_tube}
\par The basic design concept allows flexible creation of polymer crystals. It functions by projecting wave energy into the device cavity chamber which in turn contain trapped standing waves. Here the pressure waves interact with the liquid phase monomer/cross-linker. The resulting acoustic field depends on constructive and destructive interference of the acoustic waves which depend on sample properties and separation of the transducer and reflector. The distance is tuned to be a multiple of half wavelength to form the standing wave. Hence this design allows pathlength phase adjustment between the transducer and the reflector, which is essential for fabricating high quality polymer crystals. 
\begin{figure}[ht]
\centering
\includegraphics[width=\linewidth]{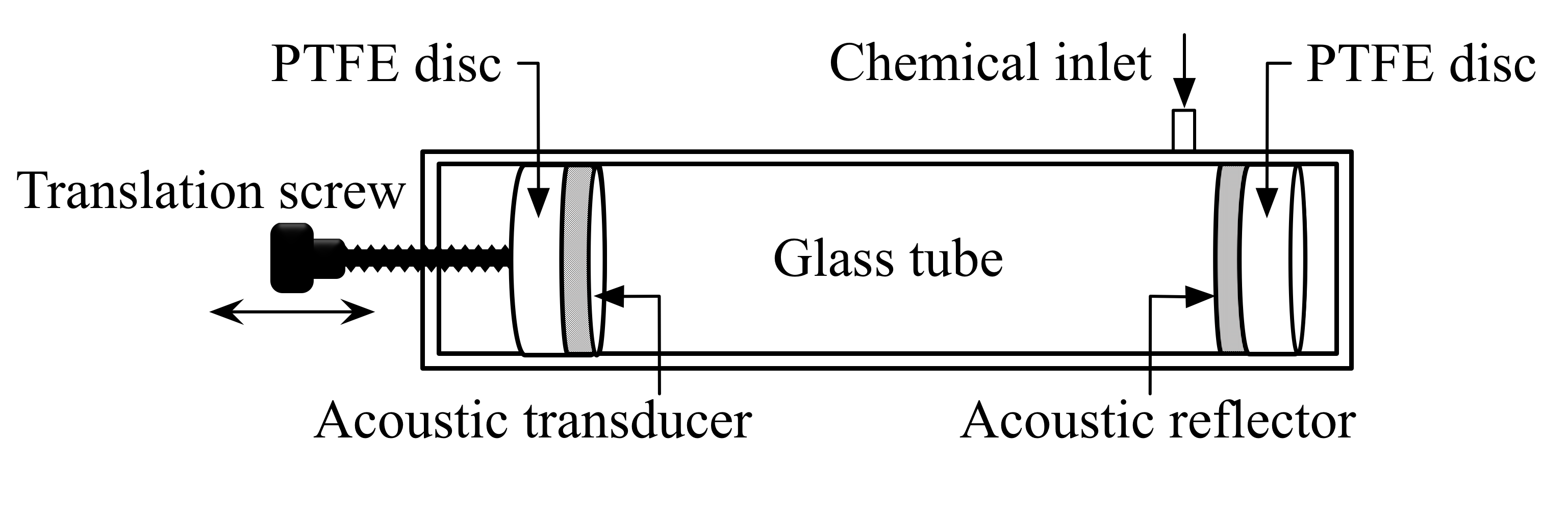}
\caption{Phononic crystal fabrication tube. The acoustic waves arise from the acoustic transducer and reflect at the boundary between the medium and the acoustic reflector. The incident wave interacts with the reflected wave constructing an acoustic standing wave. The distance between the acoustic transducer and reflector can be adjusted by turning the translation screw attached to the left PTFE disc. The chemicals are fed into the glass tube through the chemical inlet on top of the tube.}
\label{fig:PC_tube}
\end{figure}
\par The main device body consists of a glass tube (Soham Scientific, Ely, UK) with an inner diameter of 30 mm. The space available for the monomer solution is confined within the glass tube by two circular discs made from polytetrafluoroethylene (PTFE). One incorporates a 2 MHz acoustic transducer of 25mm in diameter (Noliac, Denmark) whilst the other supports an acoustic reflector made from stainless steel. The distance between the transducer and the steel reflector can be regulated by a translation screw that converts rotary motion into the linear disc motion inside the glass tube (Figure \ref{fig:PC_tube}).

\begin{table}[h]
\centering
\caption{Acoustic properties of selected solid materials \cite{mcmaster1986nondestructive}}
\caption*{\small This table is sorted according to the reflection coefficient ${R}$ in descending order. v$_l$ is the longitudinal acoustic wave travelling speed; $\rho$ is the density; $Z$ is the acoustic impedance.}
\resizebox{\linewidth}{!}{%
\begin{tabular}{l l l l l}
\hline
\textbf{material} & v$_l (mm/\mu s)$ & $\rho (g/cm^3)$&{\textit{Z} (\textit{MRayl})}&{\textit{R}}\\
\hline
platinum & 3.26&21.40&84.74&0.966 \\
gold & 3.24 &19.70 &62.60 &0.954 \\
stainless steel & 5.79& 7.69 &45.63 &0.937 \\
copper&5.01&8.93&41.46&0.931\\
silver &3.60 &1.60 &37.76 &0.925 \\
brass &4.70 &8.64 &37.30 &0.924 \\
titanium&8.27&5.15&27.69&0.899\\
iron cast&5.90&7.69&25.00&0.888\\
lead&2.20&11.20&24.49&0.886\\
aluminium & 6.42& 2.70& 17.33& 0.840\\
glass silica &5.90 &2.20 &13.00 &0.796 \\
polystyrene&2.34&1.04&2.47&0.251\\
\hline
\end{tabular}
}
\label{t:R_coefficient}
\end{table}

\par To maximise the reflected acoustic waves so that energy is largely trapped in the tube, the acoustic reflector is selected to maximise the acoustic impedance mismatch between medium and reflector. The acoustic impedance of the monomer solution is assumed to equal water at 1.48 MRayl at 20$^{\circ}$C \cite{bradley1966acoustic}. The  properties of potential materials are listed in Table \ref{t:R_coefficient}. They are sorted in descending order according to their reflection coefficient $R$. Platinum or gold produces the best reflection performance, however for this application the stainless steel disc is more accessible and practical.

\subsection{Polymeric phononic crystal fabrication}
\label{sS:PC_fab}
\par
\begin{figure}[ht]
\centering
\includegraphics[width=0.9\linewidth]{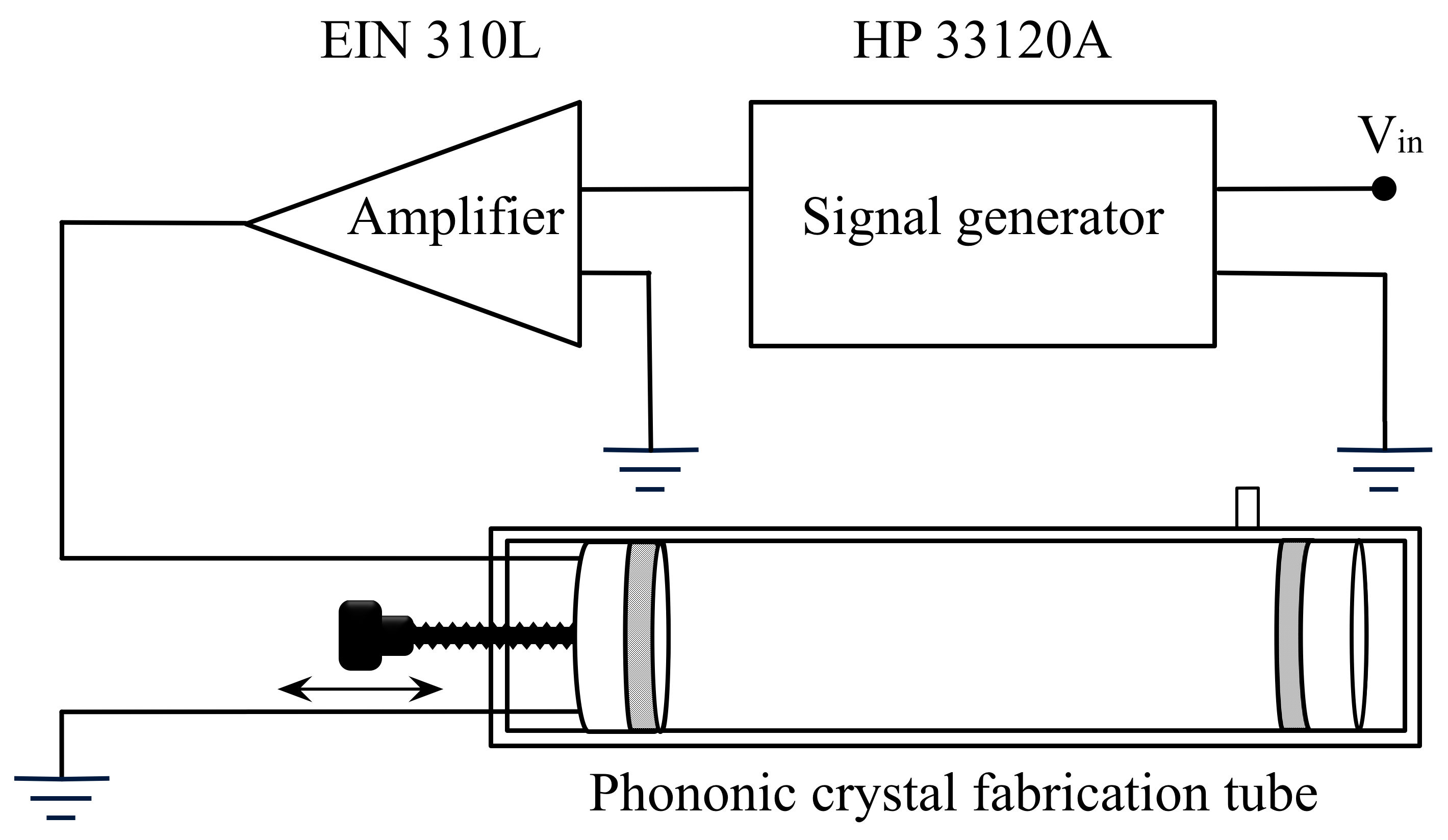}
\caption{The polymeric phononic crystal fabrication setup. The acoustic transducer in the fabrication tube is actuated by a signal generator, which passes through a power amplifier before the electric signal reaches the transducer.}
\label{fig:PC_fab}
\end{figure}

 The periodic structure that emerges is associated with the acoustic standing wave field in the fabrication tube chamber. An HP 33120A signal generator in-line with an ENI 310L power amplifier is arranged to excite the piezo disc at its natural resonant frequency. Figure \ref{fig:PC_fab} illustrates the  fabrication setup. A \SI{4}{\milli\liter} monomer/cross-linker solution containing 98.5 mol\% acrylamide and 1.5 mol\% \textit{N,N'}-methylenebisacrylamide(MBA) is used to fill the fabrication tube. The chemical initiators, including \SI{48}{\micro\liter} 10\%(w/v) freshly prepared ammonium persulfate (APS) and \SI{4}{\micro\liter} tetramethylethylenediamine (TEMED) are loaded and the chemical inlet sealed. The tube is shaken gently to achieve thorough mixing of the reagents. The device is  left on the table in a secure horizontal position. The signal generator energises the piezo disc (Noliac Ceramics NCE51) at V$_{pp}$ 90 mV at its resonant frequency.  The translation screw knob is  adjusted and a periodic structure forms in the tube that is clearly visible to the naked eye. The distance between the piezo disc and the stainless steel reflector is approximately $\sim$6 mm. The polymerisation process takes less than 1 min. The signal generator and the amplifier are switched off at a predetermined time corresponding to the onset of polymer rigidification.

\subsection{Acoustic velocity measurement of the polymer}
\label{sS:PC_avm}

\begin{figure}[ht]
\centering
\includegraphics[width=0.9\linewidth]{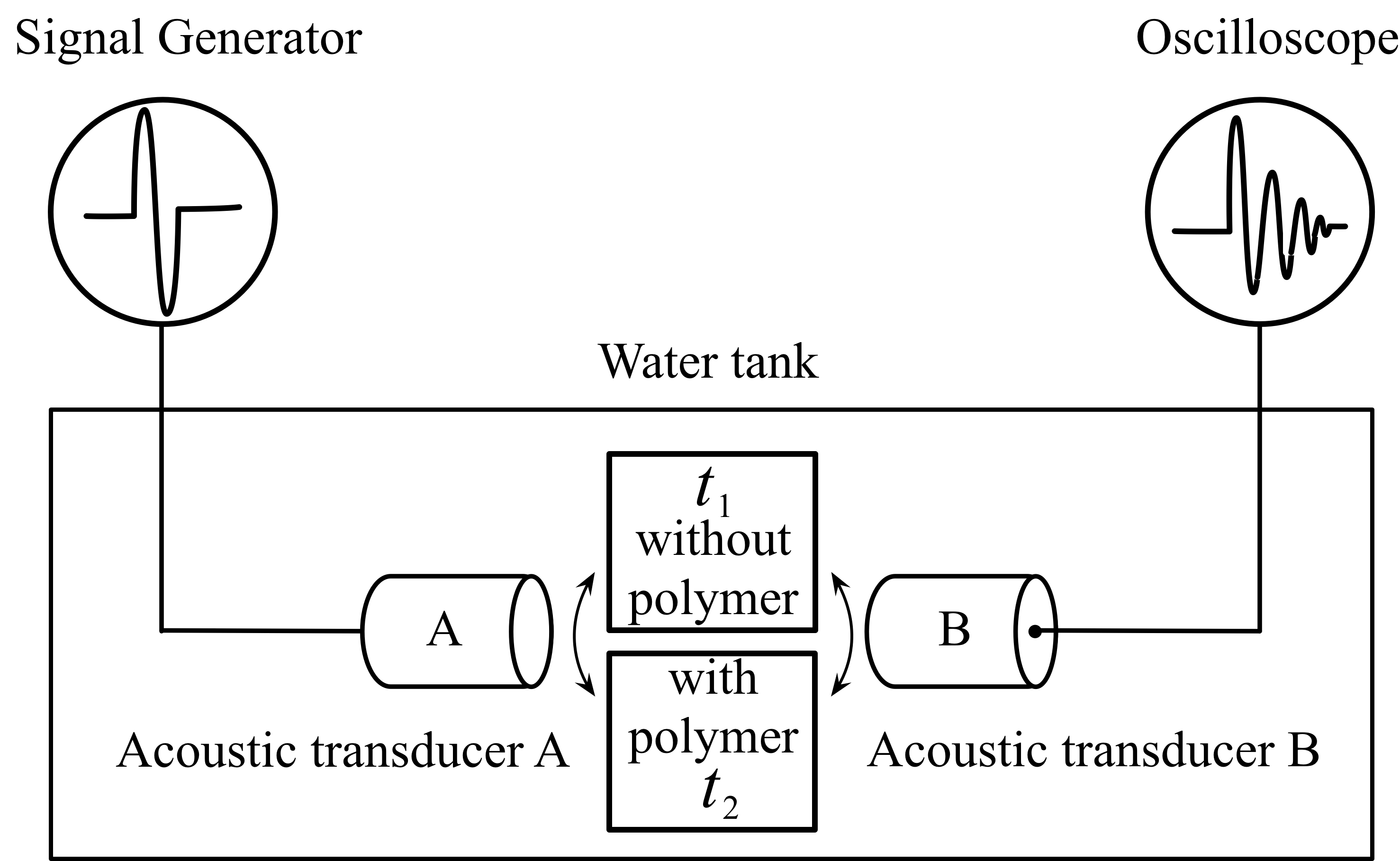}
\caption{Experimental method to determine the average acoustic velocity across the polymer. The acoustic bursts travelling time difference \(\Delta t\)=\(t_1-t_2\) is recorded, where \(t_1\) is the time that an acoustic burst travels from acoustic transducer A to B without a polymer block in between, while \(t_2\) is the travelling time when a polymer is placed between A and B.}
\label{fig:PC_avm}
\end{figure}

To investigate any slowing effect on acoustic waves we measured their average acoustic velocities and compared to their counterparts using a pulsed based  setup. Figure \ref{fig:PC_avm} illustrates how the acoustic velocity through a polymer block is measured. An acoustic signal burst generated by the transducer A travels through the tank media to reach transducer B. If a polymer block is placed in  path A-B a time difference $\Delta t$ results. Rearranging the time difference calculation Equation \ref{eq:P_avm00}, Equation \ref{eq:P_avm01} gives the average acoustic velocity through the polymer alone, $c_p$.
\begin{equation}
\Delta t=\frac{th_p}{c_p}-\frac{th_p}{c_w}
\label{eq:P_avm00}
\end{equation}
\begin{equation}
c_p=\frac{th_p}{\Delta t+th_p/c_w}
\label{eq:P_avm01}
\end{equation}
\par Where $th_p$ is the polymer thickness measured with a digital calliper. To minimise measurement error due to the polymer creep, hard plastic discs of known thickness are placed on either side of the polymer crystal during measurement. $c_w$ is the acoustic velocity in water and depends on temperature. $c_w$ used in this study is calculated from a quadratic equation $c_w$=$\sum \limits_{i=0}^{2}k_iT^i$. This simplified equation is reasonably accurate over the 15-35$^{\circ}$C temperature range \cite{lubbers1998simple}. The coefficient $k_i$ are given in Table \ref{t:avw_k} and $T$ is the temperature of water in Celsius. The water temperature used for the velocity measurements is controlled at 23$\pm$ 0.2$^{\circ}$C and corresponds to 1491 m/s.

\begin{table}[h]
\centering
\caption{Coefficients for acoustic velocity calculation in water}
\begin{tabular}{l l l}
\hline
$\bm{i}$ & $\bm{k_i}$\\
\hline
0 &  1404.30\\
1 & 4.70 \\
2 & -0.04 \\
\hline
\end{tabular}
\label{t:avw_k}
\end{table}

\section{Results and discussion}
\label{S:result}
\subsection{The polymeric crystal}
\label{sS:PC_opt}
\par The nature of the polymer crystal is related to the acoustic standing wave field and the shape of the fabrication tube.  The process employed for making the polymer cyrstal is described in section \ref{sS:PC_fab}. The resulting crystal after stabilisation is a 6mm thick polymer disc with a diameter of  30 mm.  Importantly, the disc contains distinct layers that can be seen by the naked eye. To help interpret the nature of layers we consider a simplified approach here.

\begin{figure}[ht]
\centering
\includegraphics[width=0.9\linewidth]{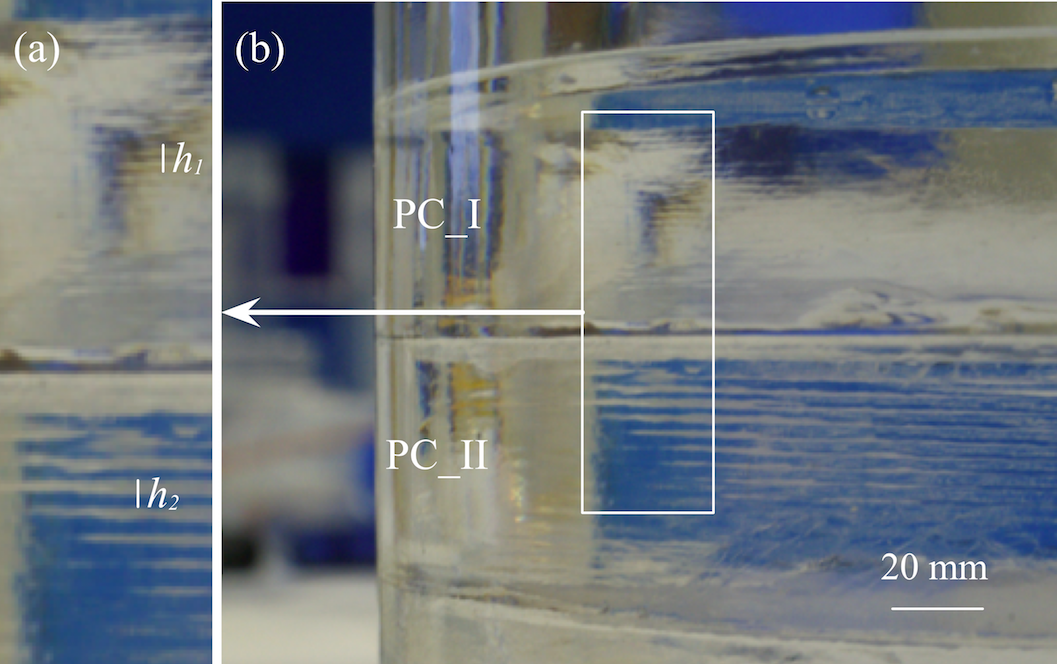}
\caption{Phononic structures made with 4 MHz (b) PC\_I and 2 MHz (b) PC\_II acoustic transducers. The labels \(h_1\) and \(h_2\) (a) are of the same length. While $h_1$covers about two spacial periodicities of the phononic crystal shown on PC\_I, $h_2$ covers only one of PC\_II.}
\label{fig:PC_comp}
\end{figure}

\par The mechanism determining the formation of the polymer crystal is still unconfirmed. Nevertheless the resulting patterns definitely emerge from pressure wave action on the monomer mix during the cross-linking process. If we move forward with a simple interpretative models, it is reasonable to conclude the rate of poylmer cross linking is proportional to the amplitude of the acoustic standing wave. Thus subject to the nodal or antinodal conditions the resulting material is likely to be biphasic, ie containing two different mechanical phases. Here the antinodes should produce a stiffer material and conversely at the nodes a less stiff material. Thus light passing through the polymer may experience a difference in propagation path leading to a visible contrast, which matches our  experimental observations.

\par
\begin{figure}[ht]
\centering
\includegraphics[width=0.9\linewidth]{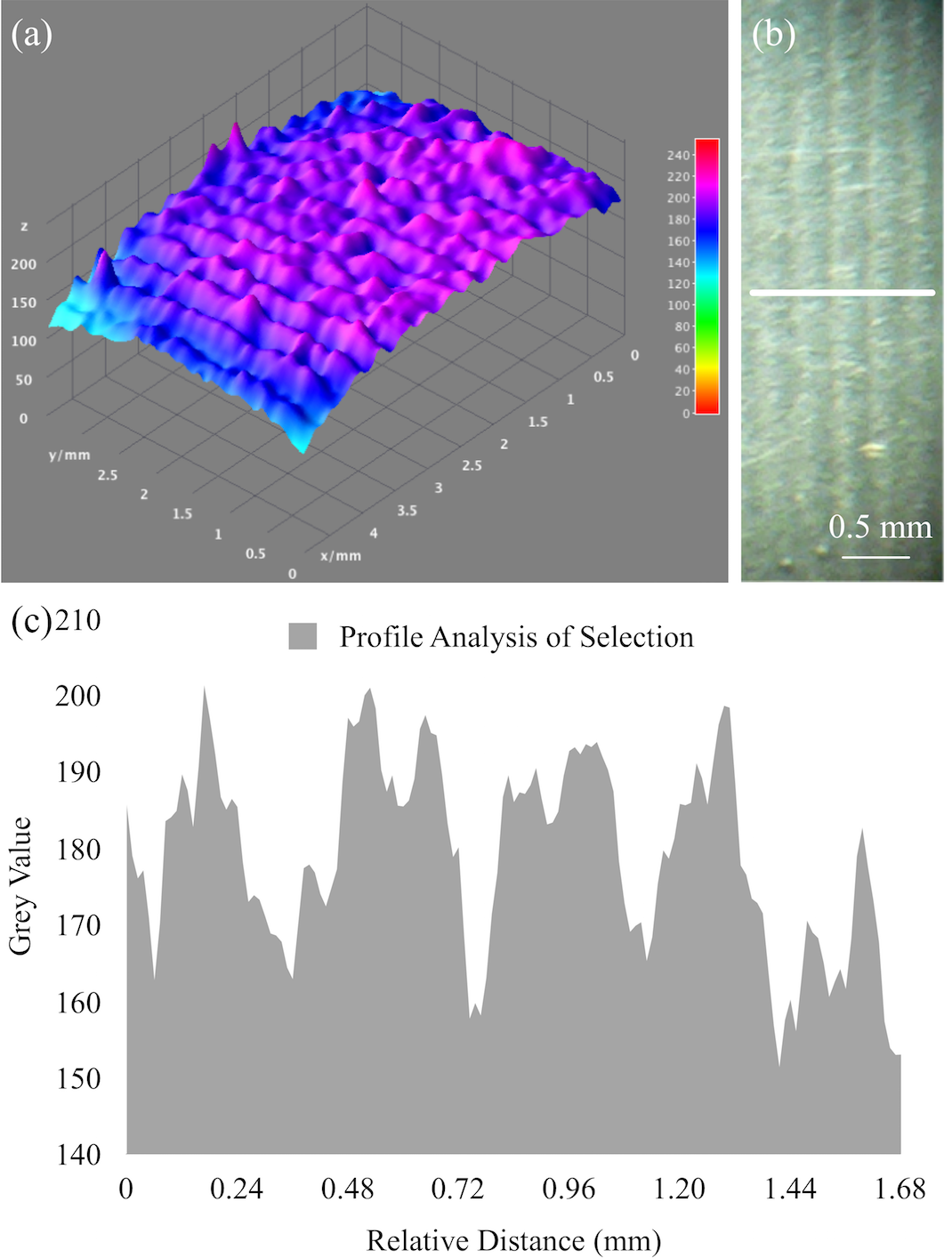}
\caption{Optical images of a  high quality polymer crystal: (a) is a simulated 3D profile via the 3D spectrum plot. Here the $z$-$axis$ shows the grey value where 0 represents black and 255 is white, and the $x$-$axis$ is perpendicular to the direction of the periodic structure and $y$-$axis$ is parallel.  (b) shows the original selected region that formed the  presented in (c)}
\label{fig:PC_slice}
\end{figure}

 Optical microscopy was used to evaluate the quality of the crystals formed. A 1 mm thick piece of polymer crystal was used to assess line definition and spacing. This was sliced perpendicular to the radial plane with the cross-section facing the lens of a 4 $times$ magnification microscope. The grey values of the selected cross-section area is plotted as a 3D surface shown in Figure \ref{fig:PC_slice} (a). To assist with periodicity analysis, a profile analysis was performed and shown in Figure \ref{fig:PC_slice} (b). The pixel grey values along a line are presented in Figure \ref{fig:PC_slice} (c) $z$-$axis$. Two light and shaded striations stand out signifying a biphasic material form.  Based on an average wavelength of  0.712$\pm$0.0077 mm (8 measurements) and a 2.16 MHz standing wave frequency the average acoustic velocity was found to be 1538 m/s, which is consistent with interpolated values for acoustic velocity \textit{vs.} acrylamide concentration in Section \ref{sS:PC_aav}. Here the patterns were coherent and consistently well-formed.

\par The transducer frequency determines the spacing in the crystal structure. The data shown in Figure \ref{fig:PC_comp} reveals the patterns can be formed with the periodicity decreasing proportionally with frequency. This is evidenced by the factor 2 increase in frequency yielding the expected factor 2 reduction in periodicity. Thus the microscopy data reveals contrast that is linked to material changes, so alternating bright and dark regions are a record of the acoustic field  during crosslinking. However it is unproven whether the fluctuations in brightness are due to a mechanical periodicity. 

\subsection{The structurised polymer under scanning electron microscope (SEM)}
\label{sS:PC_SEM}
\par
\begin{figure}[ht]
\centering
\includegraphics[width=0.9\linewidth]{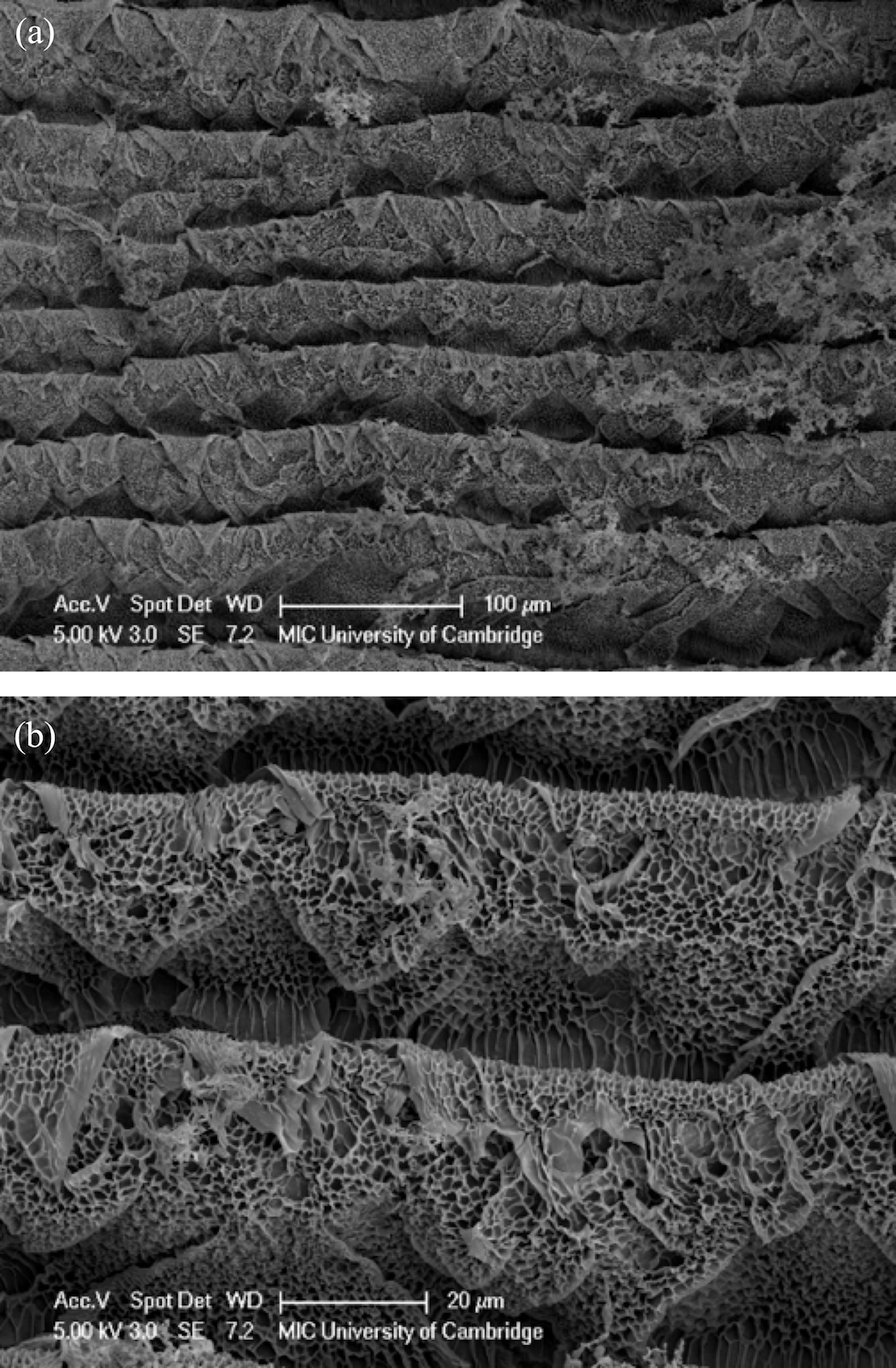}
\caption{The SEM images taken of the polymer at two different magnifications confirm that the acoustic standing wave records periodic mechanical variations in the monomer mix.}
\label{fig:PC_SEM}
\end{figure}

 SEM data was used to confirm mechanical periodicity in the polymer crystal. Here the SEM data in Figure \ref{fig:PC_SEM} shows the cross-section topology is periodic indicating an underlying variation of ‘hills and valleys’ across the sample cross-section. Furthermore these regions include elongated voids and circular voids, confirming the recording of entrenched periodic mechanical properties by the standing wave. Thus it would imply the inherent potential to scatter an acoustic wave.
\par Freeze drying of the acrylamide polymers was found to reduce structural coherence by imposing unequal forces on the structure. Thus the SEM images which were useful in identifying the mechanical periodicity of polymer, also confirms that drying of the polymer will distort the structure. Thus water content does influence the coherence of these polymer crystals.
As crystallisation patterns have been successfully recorded and confirmed the optical and SEM data it is useful to study acoustic wave transmission through them. In particular based on the theory of "the slow wave effect" we investigate the important scattering and velocity relationship.

\subsection{Average acoustic velocity across the polymer}
\label{sS:PC_aav}
\par We investigated the relation between the acoustic velocities of polymer crystals according to different polyacrylamide concentrations. As a reference point the signal from the counterpart material was also measured. These counterparts are fabricated in the same cavity device and corresponding acrylamide solutions, but during polymerisation the acoustic standing wave field is absent.  Therefore no periodic structure emerges. The polymer explore acrylamide concentrations between 5 \%(w/v) and 40 \%(w/v) in 5 \%(w/v) intervals.
\par
\begin{figure}[ht]
\centering
\includegraphics[width=\linewidth]{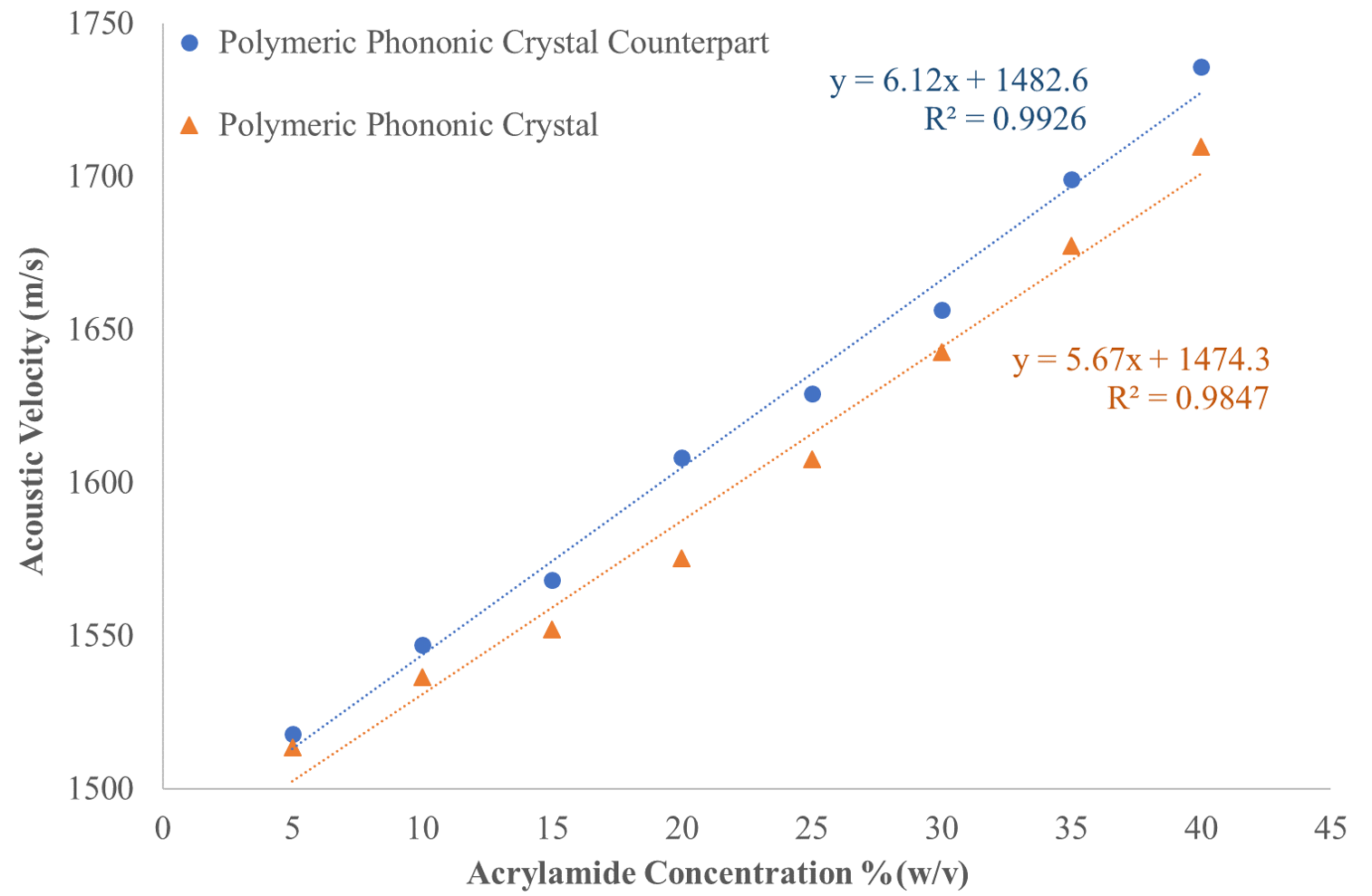}
\caption{Acoustic velocities across the polyacrylamide based phononic crystals and their counterparts made from a series of concentration monomer solutions.}
\label{fig:PC_av}
\end{figure}

 The corresponding polymer acoustic velocities are presented in Figure \ref{fig:PC_av}. A positive linear correlation between the acrylamide concentration and average acoustic velocity is observed. The general linear model function in SPSS reveals a correlation between the concentration of acrylamide and acoustic velocity across the polymer crystal. The $p$-value of interest is 0.001. The fact that it is smaller than the predetermined significance level 0.05 suggests these two fitting lines are independent.
\par A key finding is the average acoustic velocity (at 1 MHz) of the polymer crystals is lower than their bulk polymer conterparts. This agrees with the slow wave effect. Here the periodic structure of a phononic crystal interacts with the waves. Although the temporal coherence is maintained with the incident acoustic wave, the scattering has a noticeable impact on the group velocity \cite{page1997classical}. In Section \ref{sS:PC_opt} the optical evaluation of a slice of the polymer crystal indicates the average acoustic velocity through it is 1538 m/s. The monomer/cross-linker solution used in that experiment is 1.5 M equivalent to 10.85 \%(w/v). Substitute the acrylamide factor with 10.85\%(w/v) in the orange fitting line in Figure \ref{fig:PC_av}, it gives an average acoustic velocity 1536 m/s as a result in accordance with the optical evaluation value.
\par
\begin{figure}[ht]
\centering
\includegraphics[width=0.9\linewidth]{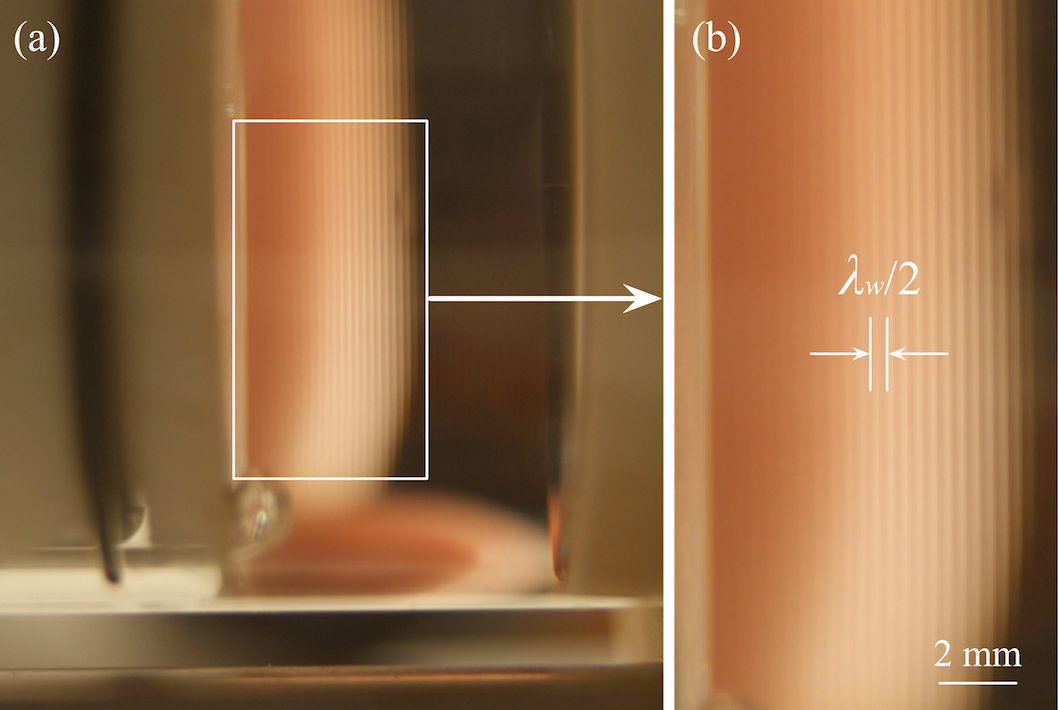}
\caption{The periodicity introduced in pure water by acoustic standing waves. $\lambda_w$ is the wavelength of acoustic wave in water and $\lambda_w$/2 labels one unit of the periodicity.}
\label{fig:PC_H2O}
\end{figure}

 One may argue that when the monomer concentration is 0 \%(w/v), the acoustic velocity $c_m$ tends to pure water $c_w$, \textit{i.e.} $\lim_{\%(w/v)\to 0}c_m=c_w$. The intercepts for both  fitting lines thus should have been set at the acoustic velocity of water at ambient temperature. However, it must be noted for pure water, the acoustic standing wave still induces a periodic structure as shown in Figure \ref{fig:PC_H2O}. Thus the surprising result is that a temporary periodic field can also lead to a velocity shift.
\par The partitioning mechanism leading to the formation of a unique periodic structure in polyacrylamide is complex. Copolymerisation of acrylamide and MBA is already more involved than the standard free radical polymerisation \cite{mcauley2004chemistry}. One perspective is that when polymerisation is taking place a sequence of events occur in the micro-environment varying pressure and energy and physico-chemical conditions over time and space. This exacerbates polymerisation complexity. Thus a working hypothesis which is one of several yet to be confirmed, is that the acoustic standing waves promote a density differential over space. As the polymer chains elongate, they become insoluble and precipitate. These precipitates have a different density from the solution. As an acoustic pressure gradient from standing wave prevails, the partially polymerised matter moves towards the nodes. Meanwhile, the monomer concentration and cross-linker also affect the polymer elasticity modulus and density \cite{bjerknes1906fields}. Overall more investigation of the event sequence is needed to isolate the mechanism.

\section{Conclusion}
\label{S:con}
\par A novel approach for fabricating polymeric phononic crystals with acoustic standing waves has been demonstrated. The cavity device follows a well established approach to create a standing wave field suitable for the polymer transformation. The starting form prior to cross-linking, comprises a 4 ml monomer/cross-linker solution plus polymerisation chemical initiators. The resulting form of the polymer crystal mirrors the cylindrical shape of the fabrication tube producing 30 distinctive layers. These have alternating refractive indices that are visible to naked eye.
\par The resultant polymeric phononic crystals were investigated by observing under an optical microscope and in transmission via an SEM. The latter provides detailed information on the periodic structure, the pore sizes in the polymer and the scale of the change. Visualising the periodic structure and determining its periodicity, provides an approximate calculation of the average acoustic velocity across the phononic crystal. The value 1538 m/s closely agrees with the interpolated value from the direct acoustic velocity measurements 1536 m/s. The refractive index differences can be attributed to variable monomer concentration affecting the elasticity. The acoustic velocity of pulses travelling through these phononic crystals and their counterparts are evaluated. The group velocity of the phononic crystals shows a lower acoustic velocity relative to counterparts of the same shape that omit periodic structures. These measurements match with theoretical studies of the behaviour of the slow wave effect in phononic crystals.
\par Contrasting with existing work this rapid fabrication approach does not necessitate the inclusion of particles, reducing complexity and saving fabrication time and costs. In its native form it is capable of creating unit cells from the standing wave planes adding important local resonance behaviour to the polymer. Thus it provides a significant shortcut for fabricating polymeric phononic crystals at both large and small scales within the time scale of one minute.

\section{Acknowledgement}
\label{S:ack}
The authors are grateful to Dr Jeremy Skepper for assistance with SEM imaging of the sample and to Dr Ke Xu Zhou for helpful discussions on the mechanism. For part of this work, Dr Nan Li was funded by the Cambridge Overseas Trust.

\section{References}
\label{S:ref}



\bibliography{bib-pc}

\begin{thebibliography}{10}

\bibitem{benchabane2006evidence}
Sarah Benchabane, Abdelkrim Khelif, J-Y Rauch, Laurent Robert, and Vincent
  Laude.
\newblock Evidence for complete surface wave band gap in a piezoelectric
  phononic crystal.
\newblock {\em Physical Review E}, 73(6):065601, 2006.

\bibitem{bjerknes1906fields}
Vilhelm Bjerknes.
\newblock {\em Fields of force: supplementary lectures, applications to
  meteorology; a course of lectures in mathematical physics delivered December
  1 to 23, 1905}.
\newblock Number~1. The Columbia university press, 1906.

\bibitem{bloch1929quantenmechanik}
Felix Bloch.
\newblock {\"U}ber die quantenmechanik der elektronen in kristallgittern.
\newblock {\em Zeitschrift f{\"u}r Physik A Hadrons and Nuclei},
  52(7):555--600, 1929.

\bibitem{bradley1966acoustic}
David~L Bradley and Wayne~D Wilson.
\newblock Acoustic impedance of sea water as a function of temperature,
  pressure and salinity.
\newblock Technical report, NAVAL ORDNANCE LAB WHITE OAK MD, 1966.

\bibitem{brillouin1953propagation}
L~Brillouin and M~Parodi.
\newblock Propagation of waves in periodic structures.
\newblock {\em Foreign literature}, 1953.

\bibitem{cheng2006observation}
Wei Cheng, Jianjun Wang, Ulrich Jonas, George Fytas, and Nikolaos Stefanou.
\newblock Observation and tuning of hypersonic bandgaps in colloidal crystals.
\newblock {\em Nature materials}, 5(10):830--836, 2006.

\bibitem{cheng2008broad}
Y~Cheng, JY~Xu, and XJ~Liu.
\newblock Broad forbidden bands in parallel-coupled locally resonant ultrasonic
  metamaterials.
\newblock {\em Applied Physics Letters}, 92(5):051913, 2008.

\bibitem{fang2006ultrasonic}
Nicholas Fang, Dongjuan Xi, Jianyi Xu, Muralidhar Ambati, Werayut
  Srituravanich, Cheng Sun, and Xiang Zhang.
\newblock Ultrasonic metamaterials with negative modulus.
\newblock {\em Nature materials}, 5(6):452--456, 2006.

\bibitem{feng2006acoustic}
Liang Feng, Xiao-Ping Liu, Ming-Hui Lu, Yan-Bin Chen, Yan-Feng Chen, Yi-Wei
  Mao, Jian Zi, Yong-Yuan Zhu, Shi-Ning Zhu, and Nai-Ben Ming.
\newblock Acoustic backward-wave negative refractions in the second band of a
  sonic crystal.
\newblock {\em Physical review letters}, 96(1):014301, 2006.

\bibitem{gillet2009atomic}
Jean-Numa Gillet, Yann Chalopin, and Sebastian Volz.
\newblock Atomic-scale three-dimensional phononic crystals with a very low
  thermal conductivity to design crystalline thermoelectric devices.
\newblock {\em Journal of Heat Transfer}, 131(4):043206, 2009.

\bibitem{hopkins2010reduction}
Patrick~E Hopkins, Charles~M Reinke, Mehmet~F Su, Roy~H Olsson~III, Eric~A
  Shaner, Zayd~C Leseman, Justin~R Serrano, Leslie~M Phinney, and Ihab El-Kady.
\newblock Reduction in the thermal conductivity of single crystalline silicon
  by phononic crystal patterning.
\newblock {\em Nano letters}, 11(1):107--112, 2010.

\bibitem{kushwaha1993acoustic}
Manvir~S Kushwaha, Peter Halevi, Leonard Dobrzynski, and Bahram
  Djafari-Rouhani.
\newblock Acoustic band structure of periodic elastic composites.
\newblock {\em Physical review letters}, 71(13):2022, 1993.

\bibitem{laude2005full}
Vincent Laude, Mika{\"e}l Wilm, Sarah Benchabane, and Abdelkrim Khelif.
\newblock Full band gap for surface acoustic waves in a piezoelectric phononic
  crystal.
\newblock {\em Physical Review E}, 71(3):036607, 2005.

\bibitem{leduc2016magnetic}
Damien Leduc, Bruno Morvan, Alain Tinel, Rebecca Sainidou, and Pascal Rembert.
\newblock Magnetic-sphere-based phononic crystals.
\newblock {\em Crystals}, 6(7):78, 2016.

\bibitem{liu2000locally}
Zhengyou Liu, Xixiang Zhang, Yiwei Mao, YY~Zhu, Zhiyu Yang, Che~Ting Chan, and
  Ping Sheng.
\newblock Locally resonant sonic materials.
\newblock {\em Science}, 289(5485):1734--1736, 2000.

\bibitem{lubbers1998simple}
J~Lubbers and R~Graaff.
\newblock A simple and accurate formula for the sound velocity in water.
\newblock {\em Ultrasound in medicine \& biology}, 24(7):1065--1068, 1998.

\bibitem{lucklum2015rapid}
F~Lucklum and MJ~Vellekoop.
\newblock Rapid prototyping of 3d phononic crystals using high-resolution
  stereolithography fabrication.
\newblock {\em Procedia Engineering}, 120:1095--1098, 2015.

\bibitem{mcauley2004chemistry}
KB~McAuley.
\newblock The chemistry and physics of polyacrylamide gel dosimeters: why they
  do and don't work.
\newblock In {\em Journal of Physics: Conference Series}, volume~3, page~29.
  IOP Publishing, 2004.

\bibitem{mcmaster1986nondestructive}
Robert~C McMaster, Paul Mcintire, and Michael~L Mester.
\newblock Nondestructive testing handbook. {V}ol. 7: Ultrasonic testing {ASNT}.
\newblock {\em American Society for Nondestructive Testing, Inc, 4153 Arlington
  Plaza, Columbus, Ohio 43228, USA, 1986. 677}, 1986.

\bibitem{mohammadi2011chip}
Saeed Mohammadi and Ali Adibi.
\newblock On chip complex signal processing devices using coupled phononic
  crystal slab resonators and waveguides.
\newblock {\em AIP Advances}, 1(4):041903, 2011.

\bibitem{page1997classical}
John~H Page, Henry~P Schriemer, IP~Jones, Ping Sheng, and David~A Weitz.
\newblock Classical wave propagation in strongly scattering media.
\newblock {\em Physica A: Statistical Mechanics and its Applications},
  241(1-2):64--71, 1997.

\bibitem{page1996group}
John~H Page, Ping Sheng, Henry~P Schriemer, I~Jones, Xiaodun Jing, and David~A
  Weitz.
\newblock Group velocity in strongly scattering media.
\newblock {\em Science}, pages 634--637, 1996.

\bibitem{qiu2006acoustic}
Chunyin Qiu and Zhengyou Liu.
\newblock Acoustic directional radiation and enhancement caused by band-edge
  states of two-dimensional phononic crystals.
\newblock {\em Applied physics letters}, 89(6):063106, 2006.

\bibitem{qiu2005directional}
Chunyin Qiu, Zhengyou Liu, Jing Shi, and CT~Chan.
\newblock Directional acoustic source based on the resonant cavity of
  two-dimensional phononic crystals.
\newblock {\em Applied Physics Letters}, 86(22):224105, 2005.

\bibitem{sainidou2006locally}
R~Sainidou, B~Djafari-Rouhani, Y~Pennec, and JO~Vasseur.
\newblock Locally resonant phononic crystals made of hollow spheres or
  cylinders.
\newblock {\em Physical Review B}, 73(2):024302, 2006.

\bibitem{sakoda2004optical}
Kazuaki Sakoda.
\newblock {\em Optical properties of photonic crystals}, volume~80.
\newblock Springer Science \& Business Media, 2004.

\bibitem{schriemer1997energy}
Henry~P Schriemer, Michael~L Cowan, John~H Page, Ping Sheng, Zhengyou Liu, and
  David~A Weitz.
\newblock Energy velocity of diffusing waves in strongly scattering media.
\newblock {\em Physical Review Letters}, 79(17):3166, 1997.

\bibitem{wu2004surface}
Tsung-Tsong Wu, Zi-Gui Huang, and S~Lin.
\newblock Surface and bulk acoustic waves in two-dimensional phononic crystal
  consisting of materials with general anisotropy.
\newblock {\em Physical review B}, 69(9):094301, 2004.

\bibitem{yang2004focusing}
Suxia Yang, John~H Page, Zhengyou Liu, Michael~L Cowan, Che~Ting Chan, and Ping
  Sheng.
\newblock Focusing of sound in a 3d phononic crystal.
\newblock {\em Physical review letters}, 93(2):024301, 2004.

\bibitem{zhang2004negative}
Xiangdong Zhang and Zhengyou Liu.
\newblock Negative refraction of acoustic waves in two-dimensional phononic
  crystals.
\newblock {\em Applied Physics Letters}, 85(2):341--343, 2004.

\end{thebibliography}
\bibliographystyle{plain}







\end{document}